# Put a Ring on It: Text Entry Performance on a Grip Ring Attached Smartphone

**Monwen Shen**

**Gulnar Rakhmetulla**

**Ahmed Sabbir Arif**

Human-Computer Interaction Group

University of California, Merced

Merced, CA, USA

mshen6@ucmerced.edu

grakhmetulla@ucmerced.edu

asarif@ucmerced.edu

## Abstract

This paper presents results of a study investing effects of grip rings on text entry. Results revealed that grip rings do not affect text entry performance in terms of speed, accuracy, or keystrokes per character. It then reflects on future research directions based on the results and observations from the study. The purpose of this work is to stress the necessity of classifying and evaluating low-cost mobile phone accessories.

## ACM Classification Keywords

H.5.2 User Interfaces (D.2.2, H.1.2, I.3.6): Input devices and strategies (e.g., mouse, touchscreen); H.1.2. Models and Principles: User/Machine Systems.

## Introduction

Low-cost mobile phone accessories, such as portable power banks, panoramic pods, selfie sticks, different types of lenses for photography, screen magnifiers, VR headsets, cases, grip rings, PopSockets, etc., are becoming increasingly popular among mobile users [6], likely due to their affordability. With mobile phones, these accessories are also becoming ubiquitous. Hence, it is essential that these devices are designed with due

6

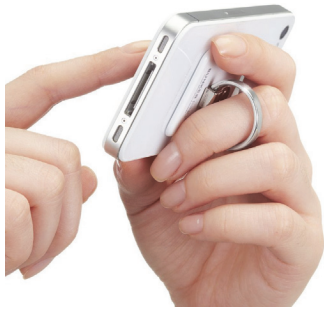

**Figure 1.** A user interacting with a grip ring attached mobile phone. From Amazon.

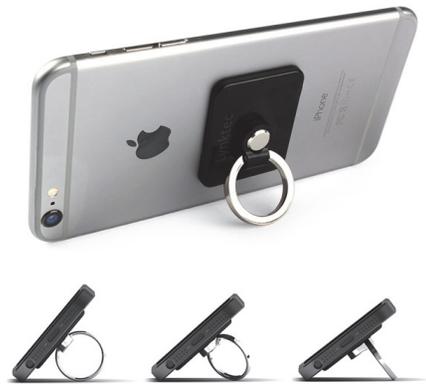

**Figure 2.** Grip ring swiveled in different degrees for different viewing angles. From Lynktec and Rakuten Global Market.

consideration of human factors. Since not much work has focused on this, we review and evaluate the effectiveness and usability of different mobile phone accessories. This exploration started with commercial selfie sticks [1] and then moved towards other low-cost accessories. This paper presents results of a user study that investigated impact of grip ring on mobile phone text entry performance in a stationary position. This is a continuation of a prior work that explored pointing accuracy on grip ring attached mobile phones [4].

## Related Work

Commonly available mobile phone grip rings enable users to hold their devices safely. Users can slip their finger through the ring to keep their mobile phone firmly in hand (Figure 1). It also serves as a stand. Users can swivel the ring up to 360 degrees to set it at the ideal video watching angle (Figure 2). Kawabata et al. [4] conducted a user study to investigate pointing accuracy on grip ring attached mobile phones. Results revealed that attaching grip ring improves pointing accuracy for smaller targets and attaching the ring at the middle yields relatively better results.

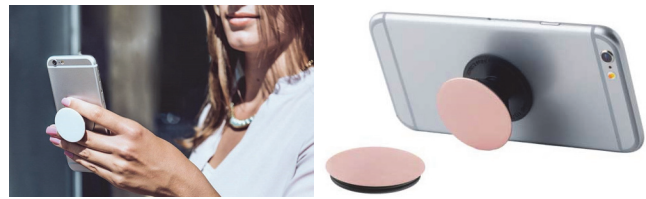

**Figure 3.** PopSocket. From Minds Alive and Glik's.

Several alternatives to grip rings are available, most popularly PopSocket [7] and LAZY-HANDS [8]. The first is a collapsible grip and stand that is expanded to use (Figure 3) and collapsed to lay flat. The second is a two to four-loop attachment that enables thumb-free grip of mobile phones (**Figure 4**). Unlike most other grips, it cannot be used as a stand.

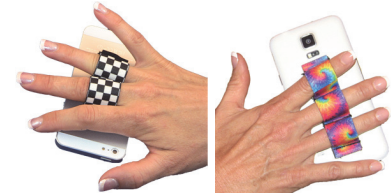

**Figure 4.** Two and three-loop LAZY-HANDS grip. From LAZY-HANDS.com

## User Study

We conducted a user study to investigate any potential impact of grip ring on text entry performance.

*Apparatus*

We used a Google Pixel XL smartphone, 154.7 × 75.7 × 8.5 mm, 168 g, at 534 ppi density during the study. We attached a Bunker Ring [9], 22.1 mm, at the center of the phone since this position yielded better pointing accuracy in a prior study [4]. Bunker Ring is one of the most sold grip rings on Amazon [9]. We used WebTEM [2] to record text entry performance.

*Participants*

Seven participants, three female and four male, average age 27.4 years (SD = 3.8) participated in the study. All were right-handed and experienced mobile users (over six years' experience). Only one (female, 28 years) used a grip ring on her mobile device.

*Design*

We used a within-subjects design for the user study. There were two conditions: with and without grip ring. The conditions were counterbalanced. In each condition, participants transcribed 15 short English phrases [5] using WebTEM [2]. In summary the design was: 7 participants × 2 conditions × 15 phrases = 210 phrases, in total. Figure 5 shows a volunteer participating in the study.

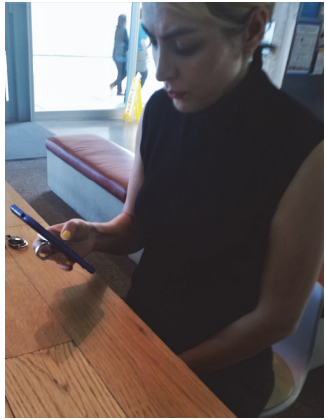

**Figure 5.** A participant transcribing text using a grip ring attached mobile phone.

*Procedure*
Upon arrival, we demonstrated the grip ring to all participants and allowed them to try it. We then started the study, where each participant transcribed 15 short English phrases from a set [5] using a smartphone with and without grip ring. WebTEM [2] displayed one random phrase from the set at a time and asked them to transcribe it. Once done, participants had to press the "Enter" key to see the next phrase. All participants transcribed text in a seated position (Figure 5). They were instructed to hold the device in portrait position with the dominant hand and then input with the thumb of the same hand. They used the default Android keyboard. However, we disabled all predictive features, including the prediction bar, auto-correction, capitalization, and custom dictionary, to eliminate a potential confound. Error correction was recommended, but was not forced [3]. WebTEM [2] logged all major text entry performance metrics [3]. At the end of the study, participants completed a short questionnaire about the grip ring.

## Results
We used a repeated-measures ANOVA for all analysis.

*Words per Minute (WPM)*
An ANOVA failed to identify a significant effect of grip ring on entry speed ($F_{1,6} = 0.05$, $p > .05$). On average entry speed without and with grip ring were 25.56 WPM (SE = 1.09) and 25.59 WPM (SE = 1.08), respectively. Figure 6 illustrates this.

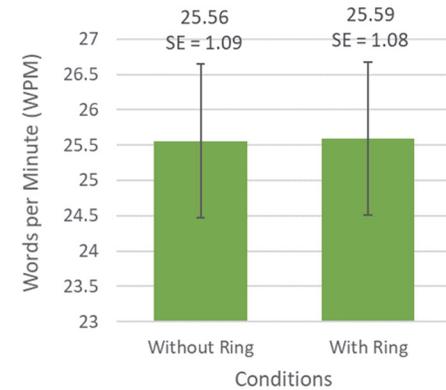

**Figure 6.** Average entry speed for the two examined conditions. Error bars represent ±1 standard error (SE).

*Error Rate (ER)*
An ANOVA failed to identify a significant effect of grip ring on ER ($F_{1,6} = 1.08$, $p > .05$). On average ER without and with grip ring were 0.32% (SE = 0.09) and 0.46% (SE = 0.09), respectively. See Figure 7.

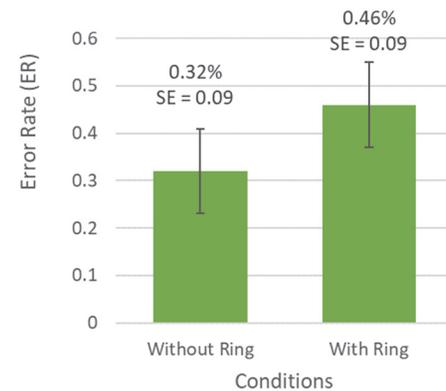

**Figure 7.** Average error rate for the two examined conditions. Error bars represent ±1 standard error (SE).

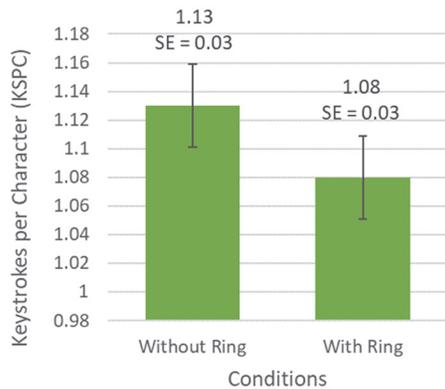

**Figure 8.** Average keystrokes per character for the two examined conditions. Error bars signify ±1 standard error (SE).

*Keystrokes per Character (KSPC)*
An ANOVA failed to identify a significant effect of grip ring on KSPC ($F_{1,6} = 2.68$, $p > .05$). On average KSPC without and with grip ring were 1.13 (SE = 0.03) and 1.08 (SE = 0.03), respectively. Figure 8 illustrates this.

*Qualitative Data*
Most participant (57%, $N = 4$) did not feel that grip ring had any impact on entry speed or accuracy, while 29% ($N = 2$) felt it improved their entry speed and accuracy. The remaining 14% ($N = 1$) were neutral. Only 29% ($N = 2$) responded that they would consider using grip ring on their mobile devices, one of them was already using a grip ring. 57% ($N = 4$) responded that they would not use grip rings, while the remaining 14% ($N = 1$) were neutral.

*Grip Finger*
We observed that participant exclusively used index finger (71%, N = 5) or the ring finger (29%, N = 2) for the grip ring. This could be due to differences in hand sizes. However, we do not have sufficient data to fully investigate this.

## Discussion
Results revealed that there was no significant effect of grip ring on text entry performance. Both conditions yielded comparable entry speed, accuracy, and keystrokes per character. However, since target selection is arguably more difficult in mobile settings, such as while walking and commuting, grip ring may benefit text entry on the go.

Qualitative data showed that most participants were reluctant on using grip rings on their devices. It may be worthwhile to investigate if there is a link between mobile phone usage and users' interest in using grip rings or similar devices. In other words, whether heavy mobile users are more likely to adapt to these devices or not. Interestingly, we observed that different users use different fingers with grip rings. This could be due to different hand sizes. Yet, we cannot validate this due to insufficient data.

## Future Work
In the future, we will include more participants in the study. Since grip rings are likely to be more useful while walking, we will investigate whether it influences text entry performance on the go. We will also explore if handedness and different hand sizes impact grip ring usage, preference, and performance. Finally, we will expand our investigation to various grip ring alternatives, such as PopSockets and LAZY-HANDS.

## Conclusion
We presented results of a study that suggested that grip rings do not affect text entry performance in terms of speed, accuracy, and keystrokes per character. We then reflected on future research directions based on the results and observations. The purpose of this work was to highlight the importance of classifying and evaluating low-cost mobile phone accessories.

## Proposed Workshop Scenario
In the workshop, we wish to discuss potential effects of various low-cost mobile phone accessories on text entry performance. Our intent is to highlight the importance of categorizing and evaluating these devices. We also wish to discuss why certain mobile phone accessories are more popular in some countries than in the others, focusing on the sociotechnical aspects of these accessories.

## Biography
This section presents short biographies of the authors.

**Monwen Shen** is a MS student at the University of California, Merced. His research interests include embedded systems and Internet of things.

**Gulnar Rakhmetulla** is a PhD student at the University of California, Merced. Her research focuses on text entry evaluation metrics and the design and development of accessible text entry techniques.

**Ahmed Sabbir Arif** is an Assistant Professor at the University of California, Merced, where he leads the Human-Computer Interaction Group. A major thread of his research focuses on smarter solutions for text entry and editing. He also models text entry performance and designs text entry techniques for underrepresented languages and user groups.